# Evaluating the local bandgap across In$_x$Ga$_{1-x}$As multiple quantum wells in a metamorphic laser via low-loss EELS

Nicholas Stephen[*,1], Ivan Pinto-Huguet[2], Robert Lawrence[3], Demie Kepaptsoglou[3,4], Marc Botifoll[2], Agnieszka Gocalinska[5], Enrica Mura[5], Quentin Ramasse[4,6], Emanuele Pelucchi[5], Jordi Arbiol[2,7] and Miryam Arredondo[1*]

[1] *School of Mathematics and Physics, Queen's University Belfast, University Road, Belfast, UK.*
[2] *Catalan Institute of Nanoscience and Nanotechnology (ICN2), CISC and BIST, Campus UAB, Bellaterra, 08193 Barcelona, Catalonia, Spain.*
[3] *School of Physics, Engineering and Technology, University of York, Heslington, York, UK.*
[4] *SuperSTEM Laboratory, SciTech Daresbury Campus, Daresbury, UK.*
[5] *Tyndall National Institute, University College Cork, "Lee Maltings", Dyke Parade, Cork, Ireland.*
[6] *School of Chemical and Process Engineering and School of Physics and Astronomy, University of Leeds, Leeds, UK.*
[7] *ICREA, Pg. Lluis Companys 23, 08010 Barcelona, Catalonia, Spain.*
*\*nstephen01@qub.ac.uk, \*m.arredondo@qub.ac.uk*

## Abstract

Using high resolution scanning transmission electron microscopy and low-loss electron energy loss spectroscopy, we correlate the local bandgap ($E_g$), indium concentration, and strain distribution across multiple In$_x$Ga$_{1-x}$As quantum wells (QWs), on a GaAs substrate, within a metamorphic laser structure. Our findings reveal significant inhomogeneities, particularly near the interfaces, for both the indium and strain distribution, and subtle variations in the $E_g$ across individual QWs. The interplay between strain, composition, and $E_g$ was further explored by density functional theory simulations, indicating that variations in the $E_g$ are predominantly influenced by the indium concentration, with strain playing a minor role. The observed local inhomogeneities suggest that differences between individual QWs may affect the collective emission and performance of the final device. This study highlights the importance of spatially resolved analysis in understanding and optimising the electronic and optical properties for designing of next-generation metamorphic lasers with multiple QWs as the active region.

## Introduction

The integration of metamorphic buffers (MB) into laser structures has greatly advanced the device design and fabrication of semiconductor devices, particularly for telecommunications. Metamorphic lasers incorporate a buffer layer between the substrate and active region to mitigate lattice constant mismatches, facilitating the development of novel laser structures, in which the optoelectronic properties can be enhanced or tuned. Particularly, metamorphic In$_x$Ga$_{1-x}$As (where x is the mole fraction) quantum well (QW) lasers have demonstrated significant advantages in this regard[1-4], with multiple QWs (MQWs) offering clear benefits over single QWs, including reduced threshold current[5], decreased temperature sensitivity[6] and improved quantum efficiency[7]. Moreover, by adjusting the thickness[8, 9] and number of QWs[10], the emission wavelength can be tuned, due to carrier interactions within the laser structure[11, 12], thus enhancing laser performance. However, these performance improvements often assume a certain degree of homogeneity across QWs, a condition heavily influenced by factors such as chemical composition and strain, both governed by the growth conditions.

Central to the performance of semiconductor lasers is the bandgap energy ($E_g$), which governs optical transitions. The control and engineering of the $E_g$ is key to designing lasers for specific applications. $In_xGa_{1-x}As$ has a direct $E_g$, that is mainly linked to the Indium (In) concentration, enabling the tuning of laser emissions to wavelengths ranging from 1300nm to 1550nm, predominantly used for telecommunication applications[1, 13, 14]. In concentration also affects the lattice constant[15], which in turn determines the type and magnitude of strain within the laser's active region and adjacent layers. For example, regions with higher In content experience greater compressive strain. Strain, both compressive and tensile, can significantly modify the $E_g$ by altering the orbital energy levels[16-22]. As a result, in addition to controlling the In concentration, strain engineering is a common method for fine tuning the $E_g$. However, the impact of specific In and strain distributions within QWs is still uncertain, mainly due to the limited ability to directly correlate nanoscale variations with changes in the QWs' electronic structure, including $E_g$. Traditionally, a variety of techniques, including UV-Vis spectroscopy[23], X-ray photoelectron spectroscopy[24] and photoluminescence[25] are used to measure the $E_g$. While these are effective for bulk materials, they are often limited by the spatial resolution required to capture nanoscale phenomena that occur in thin films and other nanostructures such as QWs and quantum dots (QDs), where heterogeneity in composition, strain, and defects can create localised states within the $E_g$. In contrast, low-loss electron energy loss spectroscopy (EELS) offers a significant advantage by probing the electronic structure at high spatial resolution with comparable energy resolution to traditional techniques, making it an ideal tool for measuring the $E_g$ at the nanoscale[26-30]. This technique has been applied to a variety of semiconductor materials, including $Cu(In,Ga)Se$[31], $In_xGa_{1-x}N$ QWs[32], CdSe QDs[26], alpha-alumina grain boundaries[33], $WS_2$ nanoflowers[34] and $Mo_xW_{1-x}S_2$ nanoflakes[35]. However, spatially resolved $E_g$ measurements for $In_xGa_{1-x}As$ QWs remain scarce, with most studies focusing on nanostructures, such as InAs/InGaAs nanowires[36], and few studies examining device structures[32].

Understanding and controlling local inhomogeneity in QWs within metamorphic lasers is key to advancing technologies in telecommunications, photonics, and beyond, as these directly influence the optoelectronic properties and performance of devices. In this study, we map the In concentration and strain distribution across three stacked $In_{0.40}Ga_{0.60}As$ QWs in an MB laser structure at the nanoscale. Using low-loss EELS, we correlate these spatial variations to the $E_g$ values, providing insight into the complex nanoscale interplay between composition, strain, and bandgap within the QWs. Our findings suggest that variations across individual QWs may significantly impact laser emission, offering valuable insight into designing the next generation of metamorphic lasers with multiple QWs.

**Results and Discussion**

**Structure overview.** Figure 1a is an annular dark field-scanning transmission electron microscopy (ADF-STEM) image displaying a cross-sectional overview of the full metamorphic laser structure, further detailed in Fig. S1, and elsewhere[37]. The high angle annular dark field (HAADF) STEM image in Fig. 1b displays the active region of the metamorphic laser, that containing the MQWs. This region consists of, using nominal thicknesses, a 5nm GaAs interface controlling layer (CIL), a 7nm $In_{0.40}Ga_{0.60}As$ QW followed by a 20nm $In_{0.13}Ga_{0.87}As$ barrier, repeated three times. The measured thickness of all QWs is ~8.6nm (Table S1), with some visible roughness at the interfaces (Fig. 1b). For clarity in the next sections, the QW here referred as the bottom QW is that nearest to the $In_xGa_{1-x}As$ MB, while the top QW is that furthest away from MB.

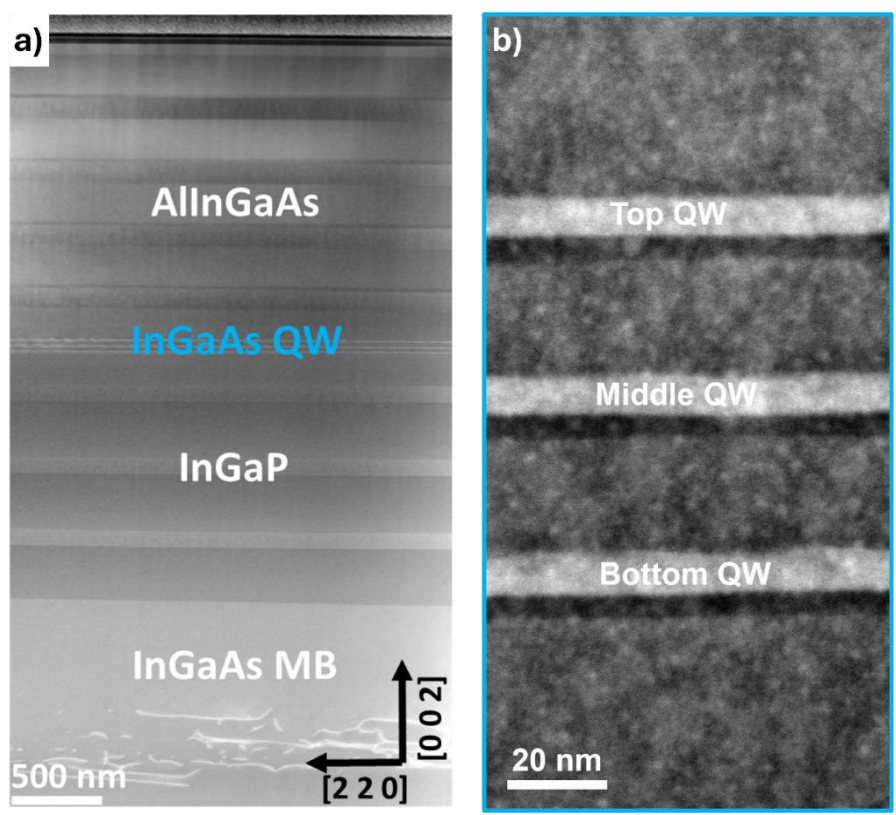

*Figure 1.* ADF-STEM sample overview (a) and HAADF-STEM of the $In_{0.40}Ga_{0.60}As$ QWs region (b).

**In atomic fraction mapping.** The relationship between indium (In) concentration and the resulting $E_g$ is well established. Increasing the In content increases the emission wavelength, thereby decreasing the $E_g$.[38, 39] For QWs, the $E_g$ is also intrinsically linked to the dimensions of the confinement, if confined. Thus, controlling the In concentration is a critical parameter in laser design and fabrication, any variations in the composition or thickness of a material such as $In_xGa_{1-x}As$ can alter the band structure, consequently affecting its $E_g$.

Figure 2 shows the In atomic fraction (at. %) maps across all QWs, acquired by energy dispersive X-ray spectroscopy (EDX) STEM (see Methods). In mole fraction, the QWs' composition is expressed as $In_{0.40}Ga_{0.60}As$, indicating that 50% of the atoms (at. %) are As, 30% are Ga and 20% are In. The relationship between mole fraction and atomic fraction is further detailed in the SI. The average In concentration for the bottom, middle and top QWs are 19.9±4.2 at. %, 21.0±4.1 at. % and 19.7±4.4 at. %, respectively, in agreement with their nominal composition. However, the distribution of In within each QW is not uniform. A significant chemical gradient is observed near the interfaces, for example reaching values as low as ~15 at. % near the interfaces to ~ 22 - 35 at. % at their centre (Fig. 2 and Fig. S2). Similar inhomogeneity has been reported for structures containing In: $In_xGa_{1-x}N$[32, 40] and $In_xGa_{1-x}As$[36, 41]. These variations are significant, as compositional asymmetry has been shown to induce shifts in the $E_g$[29, 42, 43]. An In concentration of ~30 at. % corresponds to a composition of $In_{0.60}Ga_{0.40}As$ and would result in an $E_g$ of ~0.68eV. This would represent a decrease of ~0.2eV compared to the expected $E_g$ of 0.894eV for $In_{0.40}Ga_{0.60}As$[44].

It is worth noting that the chemical composition appears less controllable in lasers grown on metamorphic buffers, likely due to the strain management required on what are overall non-planar surfaces. The surface displays a so-called crosshatched pattern, which may enhance In diffusion during growth[37]. As previously mentioned, variations in the In concentration across the QWs are expected to further impact strain levels, which modifies the lattice parameter and potentially the $E_g$. Given the strong interplay between composition and strain, we next investigate how the measured indium content affected the strain distribution within the QWs.

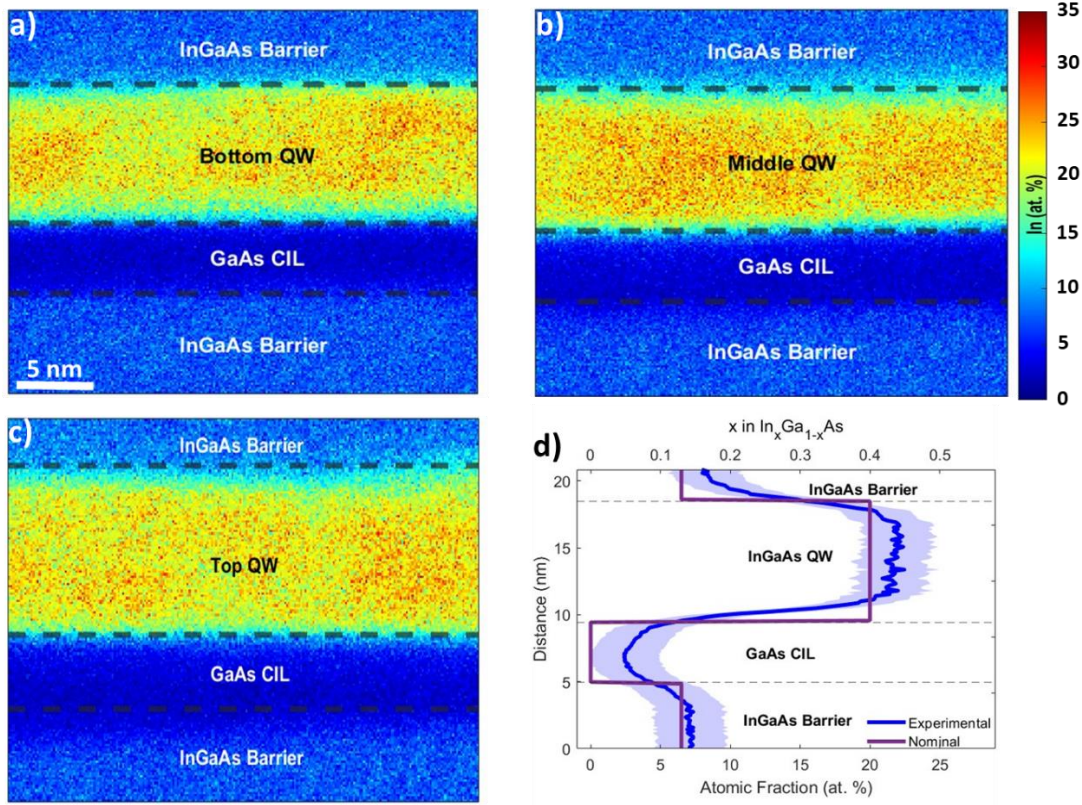

*Figure 2. In at. % colour map for the bottom (a), middle (b) and top (c) QWs region, and representative In at. % profile for the top QW (d). The blue solid line denotes the experimentally measured In at. % with its uncertainty (shaded area). The secondary x axis indicates x in $In_xGa_{1-x}As$. The pixel size is ~0.13 nm.*

**Strain Distribution.** Strain plays a critical role in modifying the electronic structure of III-V semiconductors, directly influencing the $E_g$ by altering orbital energy levels, demonstrated both theoretically[20, 45, 46] and experimentally[47, 48]. Kuo *et al.* showed that biaxal tensile strain in $In_xGa_{1-x}As$ epilayers on GaAs (and $In_xGa_{1-x}P$ on InP) leads to a reduction in the $E_g$, while the opposite effect was observed for compressive strain[16]. Similarly, Gal *et al.* experimentally demonstrated that for $In_{0.17}Ga_{0.83}As$ QWs the magnitude and sign of strain can be modified as a function of thickness, inducing changes in the $E_g$[21]. In fact, for certain combinations of strain and QW thickness, the quantum confinement has been demonstrated to further affect the band structure[18].

In this study, the relative strain along the growth direction ([200], $\varepsilon_{yy}$) of the QWs was mapped using geometrical phase analysis (GPA) on high-resolution annular bright field (ABF) STEM images using **g=004** and **g=2$\bar{2}$0** to minimise strain artifacts at the interfaces[49], Fig. 3 and Fig. S3. The reason for selecting ABF-STEM images is that HAADF-STEM images seemed to be more sensitive to surface effects, e.g. surface damage.

The reference region was the $In_{0.13}Ga_{0.87}As$ barrier layer, with a lattice constant of 5.706 Å, calculated using Vegard's law and confirmed by EDX (Fig. 2d and S2). Negative strain values indicate tensile strain relative to the reference region, and positive values indicate compressive strain.

In general, the strain varies significantly along the growth direction, while the strain parallel to the interface ($[2\bar{2}0]$, $\varepsilon_{xx}$) remains largely homogenous (Fig. S4), consistent with previous GPA strain studies on similar metamorphic laser structures[50]. Considering nominal concentrations, the lattice constants for the $In_{0.40}Ga_{0.60}As$ QWs and $In_{0.13}Ga_{0.87}As$ barrier are 5.815Å and 5.705Å, respectively. Typically, the strain for a film grown on a substrate ($\varepsilon$) is expressed as:

$$\varepsilon = \frac{a_l - a_{ref}}{a_l} \times 100\% \qquad (Eq.\ 1)$$

Where $a_l$ is the lattice constant of the layer of interest and $a_{ref}$ is the lattice constant of the reference layer. Thus, as per Eq. 1, the strain in the QWs is expected to be compressive. Similarly, the GaAs CIL has a smaller lattice constant (5.653Å) than the $In_{0.13}Ga_{0.87}As$ barrier, resulting in tensile strain. The relative strain analysis confirms that the QWs are under compressive strain and the GaAs CIL layer is under tensile strain (Fig. 3 and Fig. S3).

The relative strain distribution across the QWs (Fig. 3) closely resembles the In concentration profiles (Fig. 2), reaching a maximum at the QWs' centres and decreasing near the interfaces, with strain values as high as 4% present at the centre of the top QW. This is expected, as regions with higher indium content will experience greater compressive strain. Despite this qualitative agreement, the measured strain values in the QWs deviate from the expected values based solely on the measured In concentration. For example, according to Vegard's law, a strain value ~3.8% would correspond to an In concentration of $In_{0.68}Ga_{0.32}As$, equivalent to 34 at. % In. This estimated concentration is higher than the measured In concentration at this point (22.11±2.86 at. %), and a similar deviation is observed for the other two QWs. These discrepancies suggests that other factors, such as interface effects and surface variations in the lamellae may be influencing the GPA measurements. However, despite these variations, the relative strain distribution qualitatively agrees with the In composition profile.

The next section investigates the variations in $E_g$ across the QWs and explore their relationship to both strain and indium distribution.

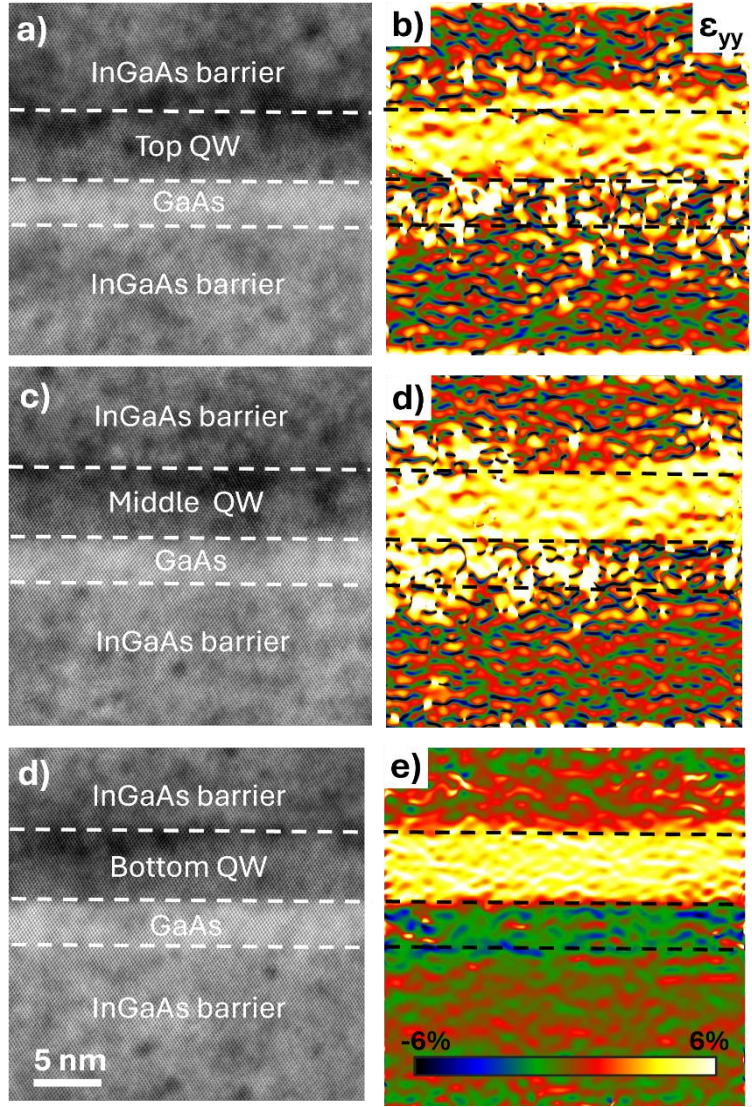

*Figure 3. Relative strain distribution across QWs. ABF-STEM viewed down [1 1 0] zone axis image (a,c,e) and corresponding $\varepsilon_{yy}$ strain map (b,d,f) for the bottom (a,b), middle (c,d) and top (e,f) QW. Pixel size is 0.039 nm.*

**Bandgap measurements across multiple QWs.** The optical $E_g$ can be phenomenologically calculated as described by Nahory et al.[44]. At 300K, the $E_g$ of $In_xGa_{1-x}As$ can be expressed in terms of the In mole fraction ($x$), as defined in Equation 2:

$$E_g(x) = 1.425\text{eV} - x\,1.501\text{eV} + x^2\,0.436\text{eV} \qquad \text{(Eq. 2)}$$

For the nominal $In_{0.40}Ga_{0.60}As$ QWs composition, the expected $E_g$ is 0.894eV. While this equation does not consider the QW thickness, known to affect the $E_g$ in $In_xGa_{1-x}As$ QWs[51], it provides a good approximation given that all QWs are nominally identical. Additionally, band bending[52] is not expected to significantly impact the QWs due to the relatively large separation between them within the laser structure, allowing each QW to be considered independently.

Low-loss EELS at 60kV and a probe size of ~1.4Å was used to measure the $E_g$ across all QWs, see Methods. The spectra were corrected for Cerenkov radiation as described in the methods and SI, following the procedure published in Ref.[27]. The error provided for each $E_g$ is the uncertainty from the fitting procedure. Cerenkov radiation occurs when electrons travel faster than the speed of light in a medium, causing a reduction in the measured $E_g$[53, 54]. Figure 4b summarises the spatially resolved and corrected $E_g$ measured across all QWs, from the bottom (position 1) to the top interface (position 6), as indicated by the yellow boxes in Fig. 4a and marked by the arrow, with a box size of 173 x 4 pixels (~15.3x0.3nm). It should be noted that the corrected $E_g$ values follow the same trend as the uncorrected data but are higher (Fig. S6). Complete $E_g$ profiles of adjacent layers are provided in Fig. S8. The average $E_g$ values for the QWs are 0.900±0.017eV for the top QW, 0.923±0.015eV for the middle QW and 0.883±0.021eV for the bottom QW (Fig. 5a). Subtle $E_g$ fluctuations within and between individual QWs are not captured when lower resolution EELS is acquired, over all QWs (Fig. S9), highlighting the importance of spatially resolved measurements.

Considering the average $E_g$ values, and their uncertainties, the highest upper value would be 0.938eV and lowest upper value 0.862, indicating a possible maximum variation of ~0.076eV between individual QWs. These values are within the range of the sample's bulk emission measured by photoluminescence (PL)[37] (Fig. S10), which has the main emission peak at ~1360nm ($E_g$ = 0.914eV) with a full width half maximum extending from 1320nm to 1410nm ($E_g$ = 0.941eV - 0.881eV). It is plausible that fluctuations in the individual $E_g$ values may contribute to the extended range observed in the PL data.

Moreover, subtle spatial variations within the individual QWs are observed, particularly for the bottom QW which exhibits an apparent asymmetry near the interfaces, with the $E_g$ values increasing towards the $In_{0.13}Ga_{0.87}As$ barrier and decreasing towards the bottom GaAs CIL.

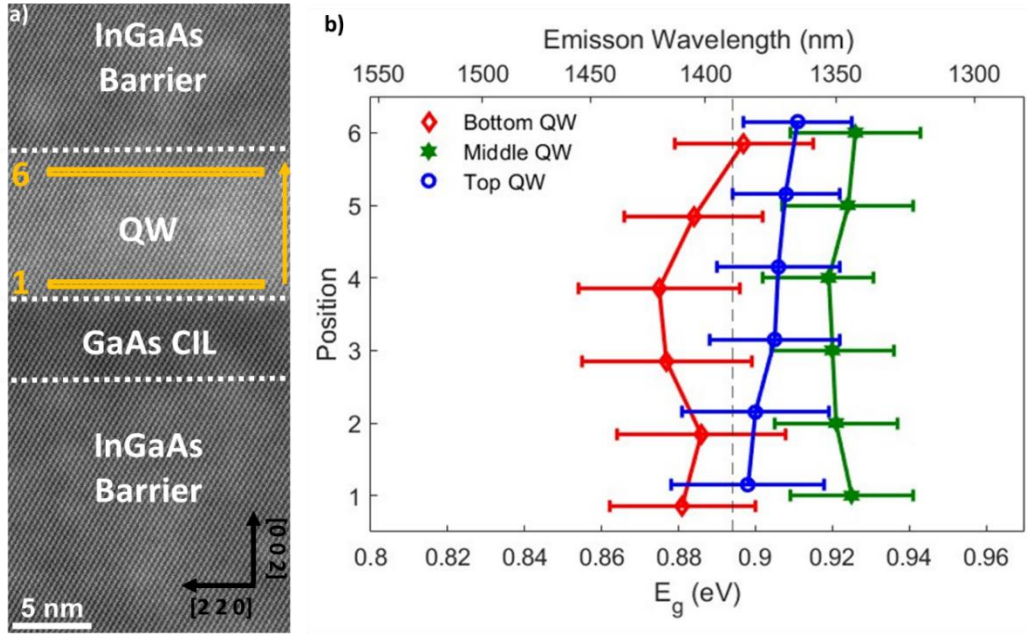

***Figure 4.*** *$E_g$ analysis. Representative HAADF-STEM image of the bottom QW (a), and the corresponding measured $E_g$ (b) corrected for Cerenkov: bottom (red diamond), middle (green hexagon) and top QW (blue circle). The $E_g$ was measured horizontally as indicated by the yellow markings in a). The secondary axis shows the corresponding emission wavelength. The dashed grey line represents the calculated $E_g$ for $In_{0.40}Ga_{0.60}As$ from Nahroy et al.[44].*

The strain is at its lowest near the interfaces, consistent with the In content, and a similar spatial variation would be expected for the $E_g$. However, surprisingly, only the bottom QW exhibits a slight but asymmetric difference in the calculated $E_g$ values near the interfaces.

Thickness variations in the areas containing the QWs were considered as a potential source of the differences in the measured $E_g$. However, this effect was neglected after carefully considering the changes in the relative thickness ($t/\lambda$) and the measured $E_g$, see Fig. S11.

To better understand the observed $E_g$ variations, we consider the composition effect. Figure 5a shows the measured average $E_g$ and In concentration for each QW. All values are near the $E_g$ calculated using Eq. 2 (dashed purple line in Fig. 5a) and within the PL measurements, demonstrating that the $E_g$ values calculated from low-loss EELS are reasonable. However, as mentioned, subtle variations in the $E_g$ values are evident across individual QWs, particularly within the bottom QW (Fig. 5b), where an asymmetry is more apparent, with only the central values falling within the expected range for the measured In concentration, considering uncertainties.

Density Functional Theory (DFT) simulations were conducted to further explore the effects of strain and composition on the $E_g$, and correlate these to the observed trends for the measured $E_g$. Figures 5c-d present the simulated low-loss EELS for various $In_xGa_{1-x}As$ alloys under different strains. These simulations consider the strain to be uniaxial (as shown in the Fig. 5d inset) and do not account for surface effects or Cerenkov radiation. Moreover, DFT underestimates the onset energy and consequently $E_g$ due to derivative discontinuity[55, 56]. Despite these limitations, this approach is known for its good reproducibility, making it ideal for identifying trends related to changes in strain and chemical composition[57, 58].

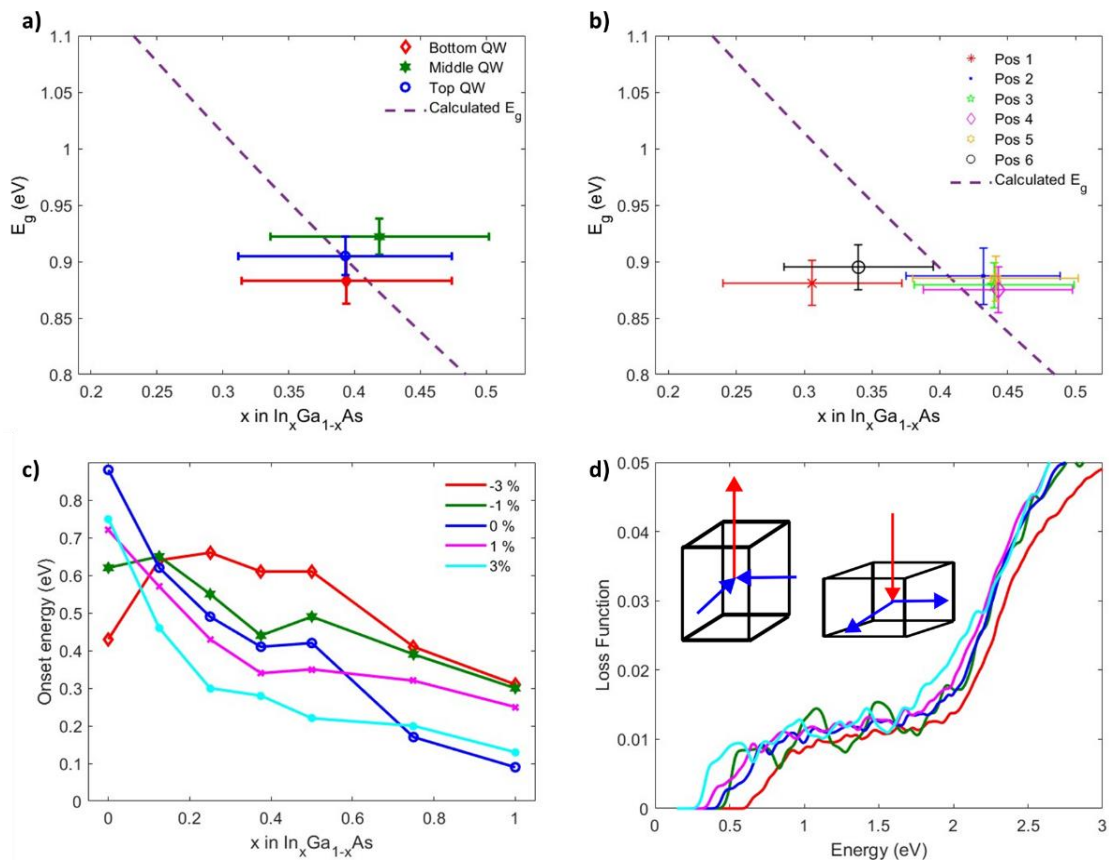

*Figure 5.* *Comparison of the calculated $E_g$ value and average measured $E_g$ values for all QWs (a) and $E_g$ values at each position (1-6) in the bottom QW (b) as function of In mole fraction (x). Calculated $E_g$ as a function of x (dotted purple line) based on Eq. 2, from Nahroy et al. [45]. Simulated EELS energy onset as a function of x in $In_xGa_{1-x}As$ (c) and for $In_{0.375}Ga_{0.625}As$ as a function of strain (d). Insert in 5d is a schematic of the uniaxial strain applied in the simulation.*

Figure 5c shows the onset energy as a function of $x$ in $In_xGa_{1-x}As$. In general, the simulations indicate that irrespective of the strain regime, the energy onset decreases as the In concentration increases, with the most drastic changes observed for the highest strain magnitudes. For compositions between $x = 0.1$ – 0.6 the compressive strain exhibits the lowest energy onset values compared to tensile strain. Figure 5d plots the change in the energy onset and the corresponding loss function for the $In_{0.375}Ga_{0.625}As$ composition, close to the QWs nominal composition (In$_{0.40}$Ga$_{0.60}$As), as a function of strain. Additional plots for other compositions are shown in Fig. S14. For this composition, the energy onset is lower for high compressive strain values and shifts to higher energies for larger tensile strain values. Further highlighting that high strain magnitudes are needed to significantly affect the E$_g$, in agreement with previous reports[46], and indicating that the In content plays a more predominant role in the resulting E$_g$ for a wide range of compositions and relatively low strain values. This provides a more direct correlation for the E$_g$ variations observed between the individual QWs and within them.

While these trends provide significant insights, they suggest there may be additional factors influencing the measured E$_g$. This opens up exciting avenues for further investigations such as interface effects, where imperfections from the growth can lead to rough or asymmetric boundaries between layers that may induce offsets which can alter the E$_g$[59, 60] and QW thickness effect[5].

## **Conclusions**

This study presents a comprehensive analysis of the spatially resolved E$_g$ chemical composition and strain across multiple In$_x$Ga$_{1-x}$As QWs within a metamorphic laser structure, employing high resolution STEM imaging and low-loss EELS techniques. Our findings reveal significant inhomogeneities in the In concentration and strain distribution across individual QWs, particularly near the interfaces.

The measured bandgap values ranged from 0.89 eV to 0.92 eV, aligning well with photoluminescence (PL) measurements but with subtle but meaningful shifts within individual QWs, critical for understanding the performance of the laser device and highlighting the utility of EELS for nanoscale E$_g$ mapping.

DFT simulations indicate that In concentration plays a dominant role in determining the local Eg, with strain having a more secondary effect, especially under low strain conditions.

These findings underscore the importance of understanding the nanoscale control over composition and strain during fabrication to optimise the optical and electronic properties of metamorphic lasers. The observed inhomogeneities could have significant implications for device performance, particularly in tuning emission wavelengths and improving efficiency. Future work should focus on refining simulation models to better account for interface imperfections and exploring methods to minimize local variations in composition and strain during growth.

## Methodology

**Sample**. The $In_{0.40}Ga_{0.60}As$ QW/GaAs metamorphic laser sample was grown via metal organic vapour phase epitaxy (MOVPE), detailed growth conditions and purity levels are reported elsewhere[37, 61]. Lamellae were prepared using a TESCAN Lyra 3 dual beam Focussed Ion Beam/Scanning Electron Microscope (FIB/SEM) via conventional *in-situ* lift out procedure[62]. Lamellae were further thinned using a Gatan Precision Ion Polishing System (PIPS) II system, followed by a final cleaning with a Fischione Model 1040 NanoMill, and baked at 125°C overnight prior to EELS acquisition. Different lamellae were used for EDX and EELS measurements.

**Electron Microscopy Techniques.** Annular dark field STEM (ADF-STEM) and energy dispersive X-ray spectroscopy (EDX) were acquired with a Thermo Fisher Talos F200-X at 200kV equipped with a four-detector in-column Super X spectrometer, offering a collection solid angle of 0.9sr, with a dwell time of 30ms and a pixel size of ~0.13 nm. The lamellae used for this were ~60 nm in thickness. For elemental quantification, Kα peaks were used to quantify Ga and As while In was quantified with the Lα peak. The background was corrected and the Brown-Powell Ionization cross section model was used in Velox[63]. In atomic fraction (at. %) maps were generated from extracting the In at. % at each pixel in the recorded EDX dataset and applying a colour map in MATLAB.

Strain mapping was carried out using the GPA plug in for Digital Micrograph from ER-C[64] on annular bright field (ABF) and HAADF-STEM images acquired on a Nion UltraSTEM 100 electron microscope operated at 100kV. The electron optics were adjusted to a convergence angle of 30mrad and a probe size of ~0.9 Å. BF-HAADF-STEM images with a pixel size of 0.039 nm were acquired as a rotational frame series (90° between frames), to eliminate stage drift and scanning distortions. The datasets were averaged by rigid and non-rigid registration[65]. ABF-STEM images were chosen for GPA analysis, using $In_{0.13}Ga_{0.87}As$ as the reference region and the **g** vectors **g=004** and **g=2$\bar{2}$0**, with a mask size of 30 pixels. Theoretical strain was calculated using the formula for strain in a layer outlined in Dunstan *et al.*[66], with lattice parameters derived by Vegard's law[15].

A Nion UltraSTEM 100MC 'HERMES', fitted with a beam monochromator and cold field emission gun, was used to perform EELS and STEM imaging[67]. The operating voltage for EELS and STEM were 60kV. The electron optics were adjusted to a convergence angle of 30mrad and a probe size of ~1.4Å. EELS spectrum images were recorded with a NION IRIS high energy resolution spectrometer equipped with a Dectris ELA hybrid pixel direct counting electron detector. The energy resolution of the experiments was ~ 30meV determined by the monochromator selection slit. The collection angle of EELS was 44mrad and spectra were acquired at 5meV/channel. All raw EELS spectra were aligned and denoised using principal component analysis (PCA) before analysis[68], as detailed in the SI.

To measure the experimental $E_g$, first, the zero-loss peak is removed from the three QWs by fitting a general power law model[69] before possible contributions from Cerenkov radiation are eliminated. Further details on this process are provided in SI. Once the spectrum is corrected, we divide the QWs into 6 horizontally equal-sized regions and obtain a global spectrum from each of them by summing the spectra of all the pixels in these regions. After preparing the spectrum and obtaining the 6 global spectra, we perform and automated fitting of the direct $E_g$:

$$I \sim A(E - E_g)^{0.5} \qquad \text{(Eq. 3)}$$

Where I is intensity of the signal, A is a constant and E is Energy[42].

**Simulations.** DFT simulations were performed using the CASTEP code[70], a plane-wave and pseudopotential based implementation of DFT. The plane wave basis set was converged to 1500eV, with a Monkhorst Pack grid of 10x10x10 points used throughout. This grid was optimised on the cell with the smallest real-space lattice parameters ensuring a minimum quality of calculation throughout. Structures were relaxed to better than an energy convergence of $2\times10^{-5}$eV per ion and a force convergence of $5\times10^{-2}$eV Å$^{-1}$, with a stress convergence of better than 1MPa. The recently developed meta-Generalized Gradient Approximation (GGA) functional RSCAN[71] was used to perform the simulations, ensuring a state-of-the-art treatment of the electron correlations within the system. This functional will, as a semi-local functional, show the derivative discontinuity problem that leads to the underestimation of the $E_g$. Finally, post-processing of the loss function was performed with the OPTADOS package[72, 73] and an adaptive smearing[74] of 0.4eV.

## **Declarations**

### **Acknowledgments**

N. Stephen and M. Arredondo acknowledge the support of the Engineering and Physical Sciences Research Council (Grant number EP/S023321/1).

E. Pelucchi acknowledges the support of Taighde Éireann – Research Ireland under Grant Numbers 15/IA/2864, 22/FFP-A/10930 and 12/RC/2276_P2.

D. Kepaptsoglou and Q. Ramasse acknowledge the support of SuperSTEM, the National Research Facility for Advanced Electron Microscopy funded by the Engineering and Physical Sciences Research Council (Grant number EP/W021080/1).

We are grateful for computational support from the UK national high performance computing service, ARCHER2, for which access was obtained via the UKCP consortium and funded by EPSRC grant ref EP/X035891/1.

ICN2 acknowledges funding from Generalitat de Catalunya 2021SGR00457. This study is part of the advanced materials programme and was supported by MCIN with funding from the European Union NextGenerationEU (PRTR-C17.I1) and Generalitat de Catalunya. We acknowledge support from CSIC Interdisciplinary Thematic Platform (PTI +) on Quantum Technologies.

**Author Contributions**

N.S and M.A conceived and planned most of the experiments. N.S and D.K carried out the EELS experiments, with D.K and Q.M.R providing support for the analysis. I.P-H, R.L, M.B and J.A provided the $In_xGa_{1-x}As$ simulations and EELS correction work. A.G, E.M and E.P provided the sample and contributed to the interpretation of the results alongside N.S and M.A. N.S took the lead in writing the manuscript. All authors provided critical feedback and helped shape the research, analysis, and manuscript.

**Conflict of Interest**

The authors declare no competing personal or financial conflicts of interest.

**Data and code availability**

Not applicable.

**Supplementary information**

Supplementary information document attached provides further details of calculations and procedures used for evaluating the $E_g$ and strain analysis. The document also provides all figures, full table of results and additional analysis that were not included as part of the main text.

**Ethical approval**

Not applicable.

**References**

[1]    I. Tångring, et al., *Journal of Crystal Growth* **2007**, 301-302, 971.
[2]    M. Arai, et al., *IEEE Journal of Selected Topics in Quantum Electronics* **2013**, 19, 1502207.
[3]    D. Wu, et al., *Electronics Letters* **2008**, 44, 474.
[4]    A. Benali, et al., *Journal of Crystal Growth* **2022**, 593, 126768.
[5]    I. Camps, et al., *Semiconductor Science and Technology* **2002**, 17, 1013.
[6]    K. Takemasa, et al., *IEEE Photonics Technology Letters* **1998**, 10, 495.
[7]    P. J. Delfyett, in *Encyclopedia of Physical Science and Technology (Third Edition)*,  (Ed: R. A. Meyers), Academic Press, New York 2003.
[8]    F. Bugge, et al., *physica status solidi c* **2006**, 3, 423.
[9]    G. Alahyarizadeh, et al., *International Journal of Modern Physics B* **2015**, 29, 1550081.
[10]   D. Saxena, et al., *Nano Letters* **2016**, 16, 5080.
[11]   P. A. Chen, et al., *IEEE Journal of Quantum Electronics* **1993**, 29, 2607.
[12]   D. S. Sizov, et al., *Applied Physics Express* **2010**, 3, 122101.
[13]   T. Uchida, et al., *Electronics Letters* **1994**, 30, 563.
[14]   A. E. Zhukov, et al., *Physics of Semiconductor Devices* **2003**, 37, 1119
[15]   I. Vurgaftman, et al., *Journal of Applied Physics* **2001**, 89, 5815.
[16]   C. P. Kuo, et al., *Journal of Applied Physics* **1985**, 57, 5428.
[17]   D. Shiri, et al., *Applied Physics Letters* **2008**, 93, 073114.
[18]   P. J. A. Thijs, et al., *Philips Journal of Research* **1995**, 49, 187.
[19]   J. Pető, et al., *npj 2D Materials and Applications* **2019**, 3, 39.
[20]   S. J. Sweeney, et al., *Journal of Applied Physics* **2019**, 125, 082538.
[21]   M. Gal, et al., *Applied Physics Letters* **1988**, 53, 113.
[22]   Y. Sun, et al., *Journal of Applied Physics* **2007**, 101, 104503.
[23]   P. Makuła, et al.,  **2018**, 9, 6814
[24]   P. V. Sushko, S. A. Chambers, *Scientific Reports* **2020**, 10, 13028.
[25]   M. A. Reshchikov, *Journal of Applied Physics* **2021**, 129, 121101.
[26]   R. Erni, N. D. Browning, *Ultramicroscopy* **2007**, 107, 267.
[27]   S. Martí-Sánchez, et al., *Nature Communications* **2022**, 13, 4089.
[28]   O. L. Krivanek, et al., *Ultramicroscopy* **2019**, 203, 60.
[29]   W. Zhan, et al., *Scientific Reports* **2018**, 8, 848.
[30]   W. Zhan, et al., *Nanotechnology* **2017**, 28, 105703.
[31]   D. Keller, et al., *Microscopy and Microanalysis* **2014**, 20, 1246.
[32]   J. R. Jinschek, et al., *Solid State Communications* **2006**, 137, 230.
[33]   J. Wei, et al., *Nano Letters* **2020**, 20, 2530.
[34]   A. Brokkelkamp, et al., *The Journal of Physical Chemistry A* **2022**, 126, 1255.
[35]   M. Pelaez-Fernandez, et al., *Nanomaterials* **2021**, 11, 3218.
[36]   D. V. Beznasyuk, et al., *Physical Review Materials* **2022**, 6, 034602.
[37]   E. E. Mura, et al., *Crystal Growth & Design* **2021**, 21, 2068.
[38]   K. H. Goetz, et al., *Journal of Applied Physics* **1983**, 54, 4543.
[39]   T. Fleischmann, et al., *Journal of Applied Physics* **2001**, 89, 4689.
[40]   T. Grieb, et al., *Ultramicroscopy* **2022**, 238, 113535.
[41]   D. V. Sridhara Rao, et al., *Microscopy and Microanalysis* **2014**, 20, 1262.
[42]   C. S. Granerød, et al., *Ultramicroscopy* **2018**, 184, 39.
[43]   C. A. Gadre, et al., *Nature* **2022**, 606, 292.
[44]   R. E. Nahory, et al., *Applied Physics Letters* **1978**, 33, 659.
[45]   G. H. Olsen, et al., *Journal of Applied Physics* **1978**, 49, 5523.
[46]   B. Mondal, R. Tonner-Zech, *Physica Scripta* **2023**, 98, 065924.
[47]   M. E. Aumer, et al., *Applied Physics Letters* **2000**, 77, 821.
[48]   J. F. Chen, et al., *Journal of Applied Physics* **2000**, 87, 1251.
[49]   J. J. P. Peters, et al., *Ultramicroscopy* **2015**, 157, 91.
[50]   N. Stephen, et al., *Journal of Materials Science* **2023**, 58, 9547.
[51]   H. A. Fonseka, et al., *Nanoscale* **2017**, 9, 13554.
[52]   D. N. Quang, N. H. Tung, *Physical Review B* **2008**, 77, 125335.

[53]   J. Park, et al., *Ultramicroscopy* **2009**, 109, 1183.
[54]   M. Horák, M. Stöger-Pollach, *Ultramicroscopy* **2015**, 157, 73.
[55]   P. Mori-Sánchez, A. J. Cohen, *Phys. Chem. Chem. Phys.* **2014**, 16, 14378.
[56]   J. P. Perdew, et al., *Physical Review Letters* **1982**, 49, 1691.
[57]   K. Lejaeghere, et al., *Science* **2016**, 351, aad3000.
[58]   Y. Ji, et al., *The Journal of Physical Chemistry A* **2022**, 126, 5924.
[59]   M. Oloumi, C. C. Matthai, *Journal of Physics: Condensed Matter* **1991**, 3, 9981.
[60]   J. Tersoff, *Physical Review B* **1984**, 30, 4874.
[61]   V. Dimastrodonato, et al., *Journal of Crystal Growth* **2011**, 315, 119.
[62]   M. Lekstrom, et al., *Journal of Physics: Conference Series* **2008**, 126, 012028.
[63]   C. J. Powell, *NIST* **1976**, 460, 97.
[64]   Geometrical phase analysis software, https://er-c.org/index.php/software/stem-data-analysis/gpa/,
[65]   L. Jones, et al., *Advanced Structural and Chemical Imaging* **2015**, 1, 8.
[66]   D. J. Dunstan, *Journal of Materials Science: Materials in Electronics* **1997**, 8, 337.
[67]   O. L. Krivanek, et al., *Journal of Physics: Conference Series* **2014**, 522, 012023.
[68]   N. Bonnet, et al., *Ultramicroscopy* **1999**, 77, 97.
[69]   K. L. Y. Fung, et al., *Ultramicroscopy* **2020**, 217, 113052.
[70]   S. J. Clark, et al., *Zeitschrift für Kristallographie - Crystalline Materials* **2005**, 220, 567.
[71]   A. P. Bartók, J. R. Yates, *The Journal of Chemical Physics* **2019**, 150, 161101.
[72]   R. J. Nicholls, et al., *Journal of Physics: Conference Series* **2012**, 371, 012062.
[73]   A. J. Morris, et al., *Computer Physics Communications* **2014**, 185, 1477.
[74]   J. R. Yates, et al., *Physical Review B* **2007**, 75, 195121.

# Supplementary Information

The full structure of the metamorphic laser consists of a GaAs (001) substrate misoriented 6° towards [1 1 1]A with a 1µm thick parabolic graded $In_xGa_{1-x}As$ metamorphic buffer (MB) with an In concentration varying nominally from $0<x<0.18$, deposited on top. After the MB, an $Al_xIn_yGa_{1-x-y}As/In_{0.62}Ga_{0.38}P$ superlattice n-type cladding layer is added with a separate confinement heterostructure (SCH) consisting of $Al_xIn_yGa_{1-x-y}As/In_{0.62}Ga_{0.38}P$ acting as the barrier layer to the active region. The active region consists of a repeated stacking of GaAs Interface Controlling layer (CIL), $In_{0.40}Ga_{0.60}As$ QW and an $In_{0.13}Ga_{0.87}As$ barrier layer as outlined in the main manuscript. After the active region, a second SCH layer is grown followed p-type $Al_xIn_yGa_{1-x-y}As/In_{0.62}Ga_{0.38}P$ SL cladding. Finally, an $In_xGa_{1-x}As$ contact layer is placed after the cladding, completing the full structure. Further details and characterisation of this structure can be found in Mura et al.[37].

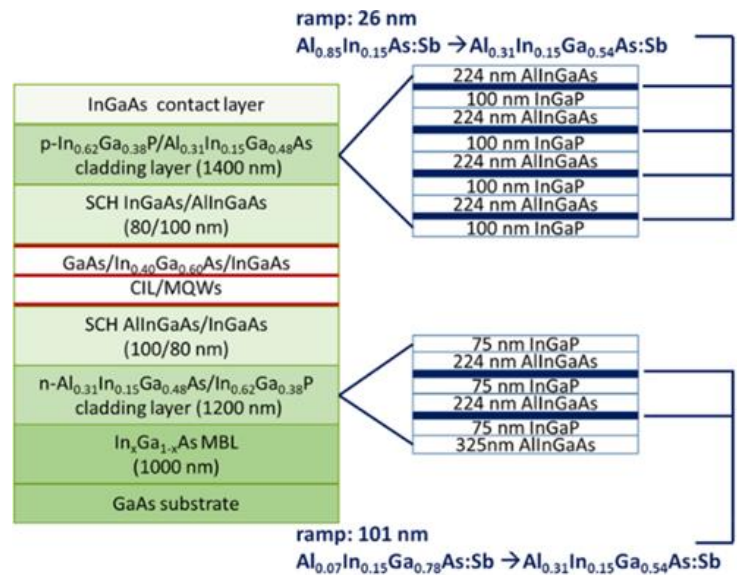

*Figure S1.* *Schematic diagram of metamorphic laser investigated. From* [1].

Given that the QW thickness can impact the bandgap ($E_g$), the thickness of each QW was measured. Table S1 shows that all QWs have comparable thicknesses, slightly thicker than the nominal 7nm.

| QW | Average Thickness (nm) |
|---|---|
| Bottom | 8.54±0.17 |
| Middle | 8.64±0.10 |
| Top | 8.60±0.17 |

**Table S1.** Measured thickness of QWs.

*In Atomic Fraction Mapping*

It should be noted that the composition $In_xGa_{1-x}As$ is expressed in mole fraction, where x is the mole fraction. This represents how much of each element (In and Ga) is in the system. The atomic fraction, presented throughout the manuscript, represents the percentage of each type of atom. Since the atomic ratio between As and the group III elements (In, Ga) is always 1:1, the As atomic fraction is always 50%. The rest of the atomic fraction is divided between In and Ga according to their mole fractions, see Table S2.

| Composition (Mole Fraction) | In Atomic Fraction (at. %) | Ga Atomic Fraction (at. %) | As Atomic Fraction (at. %) |
|---|---|---|---|
| $In_{0.20}Ga_{0.80}As$ | 10% | 40% | 50% |
| $In_{0.40}Ga_{0.60}As$ | 20% | 30% | 50% |
| $In_{0.60}Ga_{0.40}As$ | 30% | 20% | 50% |
| $In_{0.80}Ga_{0.20}As$ | 40% | 10% | 50% |

**Table S2.** Mole and atomic fraction for various $In_xGa_{1-x}As$ alloys.

Figure S2 shows the In at. % profile for both the bottom and middle QWs. Similar trends and In concentration values to those seen in the top QW (Fig. 2d) are observed. The average In at. % is presented in Table S3, indicating that the average concentrations are similar to each other.

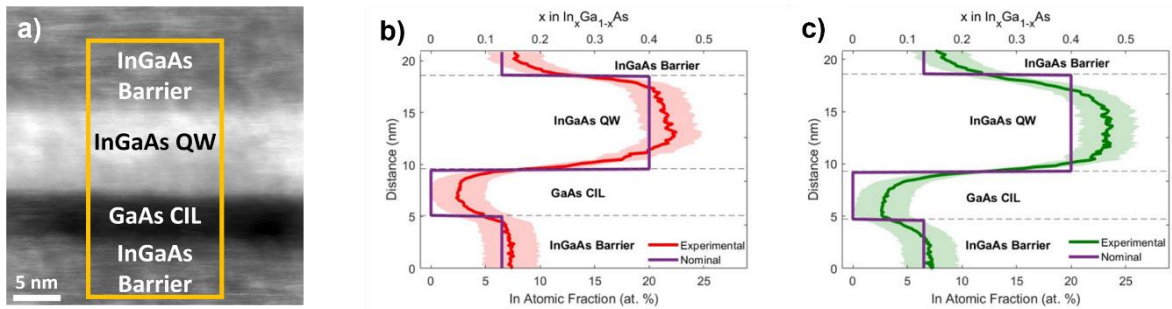

*Figure S2. Atomic fraction profile for the bottom (b) and middle QW (c), integrated over the area marked by the orange rectangle in a). The red/green solid line denote the experimentally measured In at. % and the shaded area correspond to the uncertainty in the bottom/middle QW respectively. The secondary x axis indicates to x in $In_xGa_{1-x}As$.*

| QW | In (at.%) |
|---|---|
| **Bottom** | 19.9±4.2 |
| **Middle** | 21.0±4.1 |
| **Top** | 19.7±4.4 |

*Table S3. Average In at.% across QWs.*

The concentration of the lower In$_{0.13}$Ga$_{0.87}$As barrier was also checked, used as the reference region in the Geometric Phase Analysis (GPA) and theoretical strain calculations. The In at. % in the lower In$_{0.13}$Ga$_{0.87}$As barrier for the bottom, middle and top QW are 6.8±0.9 at. %, 6.6±0.8 at. % and 6.6±0.8 at. %, respectively, as seen in Fig. S2. Thus, no significant change in the lattice parameter is expected.

**Geometrical Phase Analysis (GPA)**

As mentioned in the main manuscript, strain mapping was carried out using the GPA plug in for Digital Micrograph from ER-C[64] on annular bright field (ABF) and HAADF-STEM images acquired on a Nion UltraSTEM 100 electron microscope operated at 100kV. The electron optics were adjusted to a convergence angle of 30mrad and a probe size of ~0.9 Å. BF-HAADF-STEM images with a pixel size of 0.039 nm were acquired as a rotational frame series (90° between frames), to eliminate stage drift and scanning distortions. The dataset were averaged by rigid and non-rigid registration[65]. ABF-STEM images were chosen for GPA analysis (Fig. S3), using In$_{0.13}$Ga$_{0.87}$As as the reference region and the **g** vectors **g=004** and **g=2$\bar{2}$0**, marked as 1 and 2 respectively in Fig. S3b, with a mask size of 30 pixels, as shown in Fig. S3b. The growth direction [200] was set as ε$_{yy}$ and the direction parallel to the interface [2$\bar{2}$0] was set as ε$_{xx}$. The phase map is then tuned in the lower In$_{0.13}$Ga$_{0.87}$As barrier area. The surface blemishes observed in the HAADF and BF-STEM is likely to be surface damage. The reason for selecting ABF-STEM images is that HAADF-STEM images seemed to be more sensitive these surface blemishes which induced significant noise in the GPA processing.

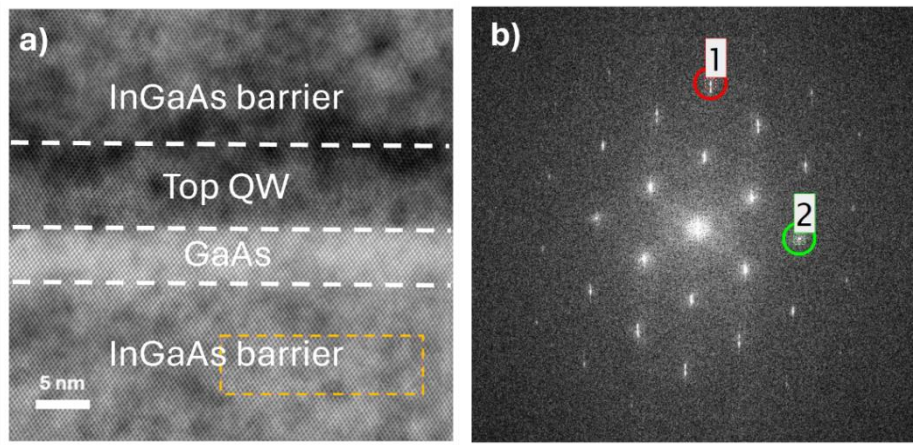

*Figure S3. Representative process of GPA strain mapping using top QW. BF-STEM image viewed down [1 1 0] zone axis (a) and corresponding FFT image (b). The GPA strain maps were obtained using the InGaAs barrier as the reference region, as highlighted by the dotted box in a), and using the **g** vectors **g=004** and **g=2$\bar{2}$0**, marked as 1 and 2, respectively. The mask size is 30 pixels. Strain mapping was carried out using the GPA plug in for Digital Micrograph from ER-C[64] on BF-STEM images with a pixel size of 0.039 nm.*

Figure S4 displays the $\varepsilon_{xx}$ and $\varepsilon_{xy}$ maps for all QWs, using the ABF-STEM images. As expected, the strain parallel to the interface ($\varepsilon_{xx}$) and the $\varepsilon_{xy}$ component is largely homogeneous. The $\varepsilon_{xx}$ maps display a large distortion on the right-hand side due to the flyback effect, a common artifact that occurs during the scanning process.

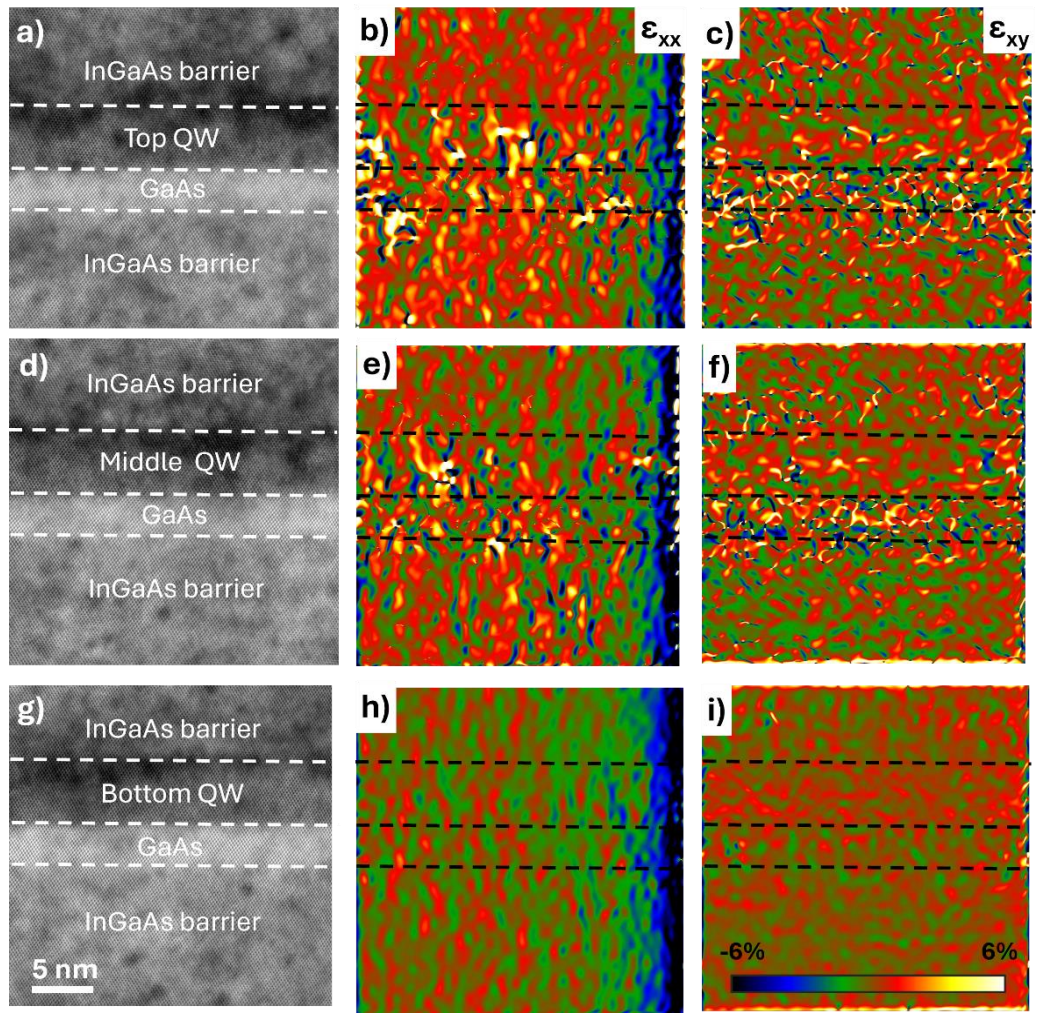

*Figure S4. GPA strain analysis for all QWs. ABF-STEM images viewed down [1 1 0] zone axis images (a,d,g). $\varepsilon_{xx}$ and $\varepsilon_{xy}$ maps for top QW (b,c), middle QW (e,f) and bottom QW (h,i). All images cropped to pixel size 750 x*

**<u>Bandgap measurements</u>**

After obtaining the raw electron energy low spectroscopy (EELS) spectra, the initial procedure involves aligning the spectra to ensure the zero-loss peak (ZLP) is precisely at 0eV. This alignment was carried out by using the "align peak" function in Digital Micrograph (DM). Examination of the ZLP (Fig. S5) post-alignment indicated that all peaks were within 0.005eV of 0eV. Considering that the resolution of the monochromator is 0.005eV, it was determined that the ZLP had been successfully aligned.

After alignment, the spectra were denoised using Principal Component Analysis (PCA) in DM. PCA determines the number of significant components present in the spectra via a scree plot, which depicts eigenvalues as function of components. The optimal number of components was determined when the eigenvalues showed minimal change *i.e.* the curve levels off. Based on the scree plots, it was consistently observed that the optimal number of significant components was 10.

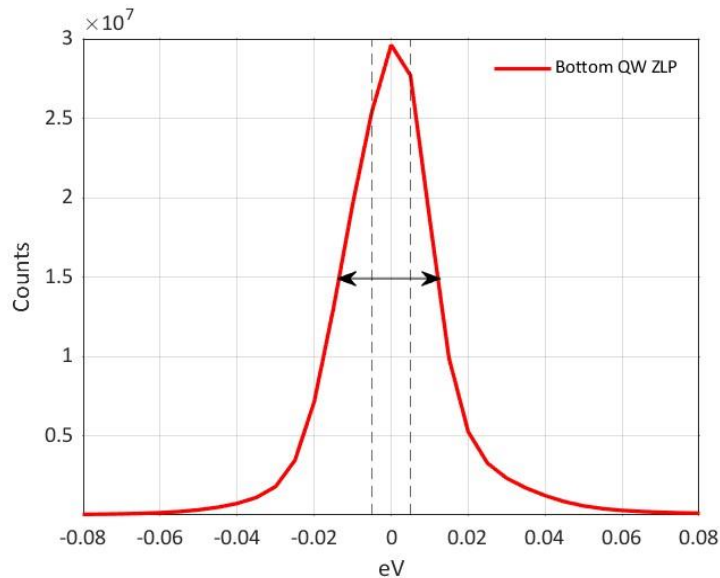

*Figure S5.* Aligned Raw EELS spectra from the bottom QW. Black dotted line denotes -0.005eV and 0.005eV. Black double head arrow represents the Full Width at Half Maxima (FWHM) for the spectra.

Due the presence of parasitic signal in the low loss, such as Cerenkov radiation, simulations are required to account for this and related effects, in order to accurately calculate the $E_g$. These simulations are based on classical electrodynamics, with the materials described by their frequency-dependent, local dielectric functions. The energy-loss probability is directly separated as the sum of a bulk contribution, which is independent of geometry and proportional to the path length traveled inside each material, and a surface term determined by the interface morphology. In these samples, the surface term is negligible due to the dielectric proximity of the different interfaces, so will only consider the bulk contribution. Both retarded ($\Gamma_{Bulk}^{R}$) and non-retarded ($\Gamma_{Bulk}^{NR}$) (*i.e.* assuming an infinite speed of light) calculations were performed and compared to assess the role of Cerenkov radiation emitted by the fast electrons in their interaction with the bulk materials. As expected, this considerably affects the region were we have to fit the $E_g$. Based on the information provided by the simulated retarded and non-retarded spectral images, we correct the experimental data by subtracting parasitic contributions associated with Cerenkov radiation. The correction function ($F_{corr}$) is the ratio between the non-retarded bulk loss and the retarded loss (Eq. S1). This function allows to transform the experimental spectrum image ($x$) to the corrected spectrum image ($F_{corr}(x)$)[27].

$$F_{corr}(x) = \frac{\Gamma_{Bulk}^{NR}}{\Gamma_{Bulk}^{R}} \cdot x \qquad \text{(Eq, S1)}$$

After that, we perform an automated bandgap fitting in this corrected spectrum. To achieve this, a fitting algorithm evaluates, for each global spectrum, all possible energy ranges within a specified range centered at the reported bulk $E_g$ value of the material. This adaptive range analysis is employed to make the computational cost manageable and to minimise the occurrence of outliers. A goodness-of-fit analysis is then conducted to determine the optimal spectral position at which these characteristic $E_g$ energy patterns are located. Given the properties of the dataset and the data analysis, the statistical parameter used for this analysis is the coefficient of determination ($r^2$). In each spectrum, we identify the energy range where the fitting yields the highest $r^2$ value and we use this fitting for calculate the $E_g$ value. The error provided for each $E_g$ is the uncertainty from the fitting procedure.

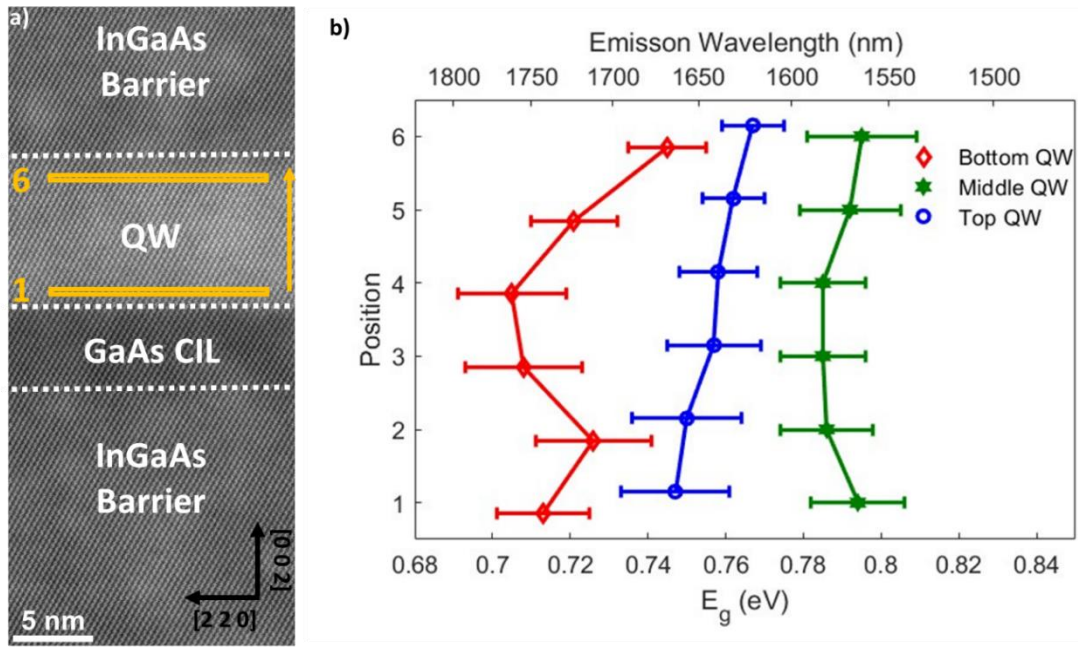

*Figure S6. $E_g$ analysis. Representative HAADF-STEM images of the middle QW (a), and the corresponding measured $E_g$ uncorrected for Cerenkov (b): bottom (red diamond), middle (green hexagon) and top QW (blue circle). The $E_g$ was measured horizontally as indicated by the yellow marking in a). The secondary axis indicates the corresponding emission wavelength.*

The gamma functions used for calculating the correction coefficient, the correction coefficient, the background spectrum image (SI) and a comparison of the original SI with the corrected one is shown in Fig. S7.

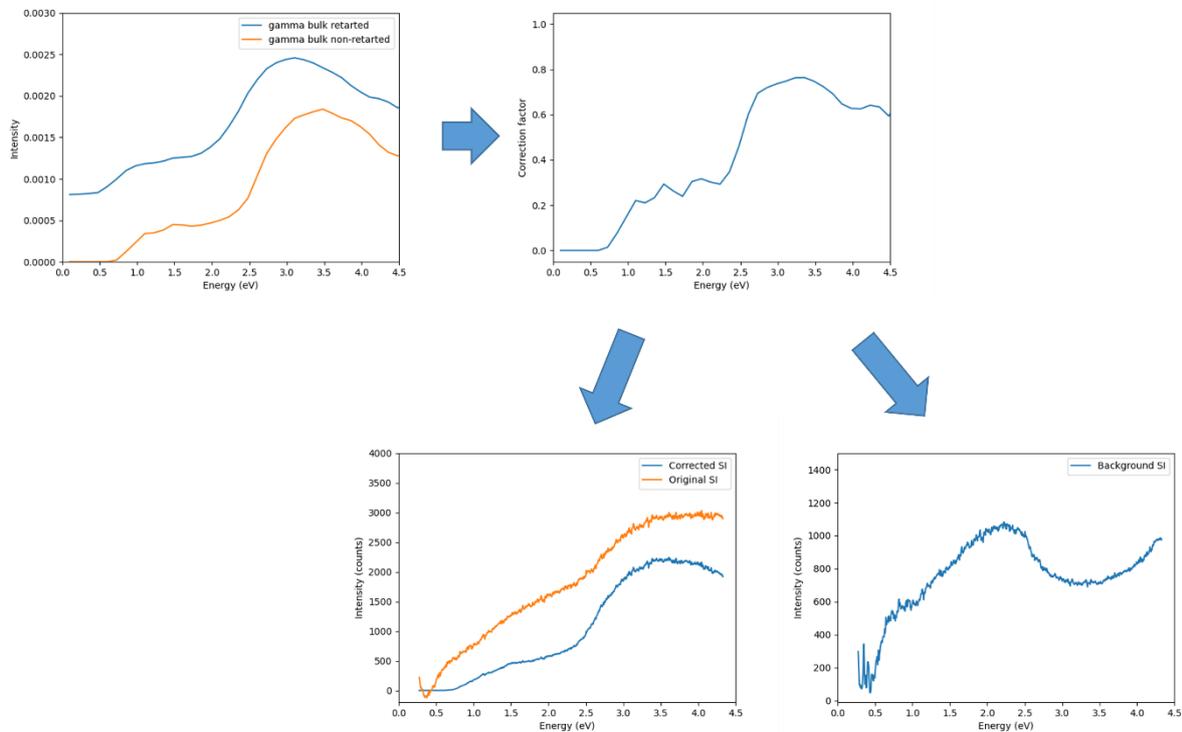

***Figure S7.*** *Bottom QW spectrum image correction. Plots of the gamma functions used for calculating the correction coefficient, the correction coefficient, the background SI and a comparison of the original SI with the corrected one.*

Figure S8 shows the complete $E_g$ profiles of adjacent layers.

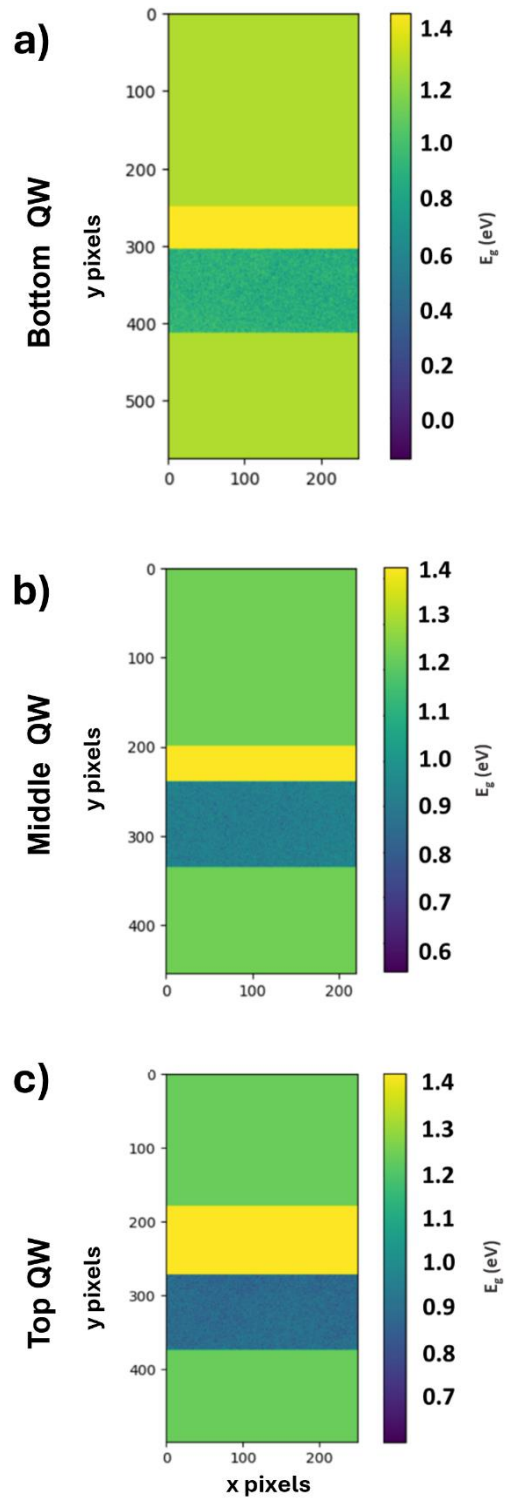

***Figure S8.*** *$E_g$ variations across adjacent layers for bottom (a), middle (b) and top (c) QW.*

Figure S9 shows the $E_g$ analysis for all the QWs in the EELS acquisition which has a lower spatial resolution compared to the higher resolution images containing the individual QW used in the main manuscript, Fig. 4. We observe that there is no significant $E_g$ variation between the position within each QW and as an average over all the QWs. This could be due the similarities in the ZLP for each QW

EELS spectra which were all nearly identical, likely because of the lower spatial resolution, that would consequently carry over when processed to calculate $E_g$. Overall, these results highlight the of the importance and motive for using individual QW EELS spectra with higher pixel resolution to resolve the $E_g$ of the QWs.

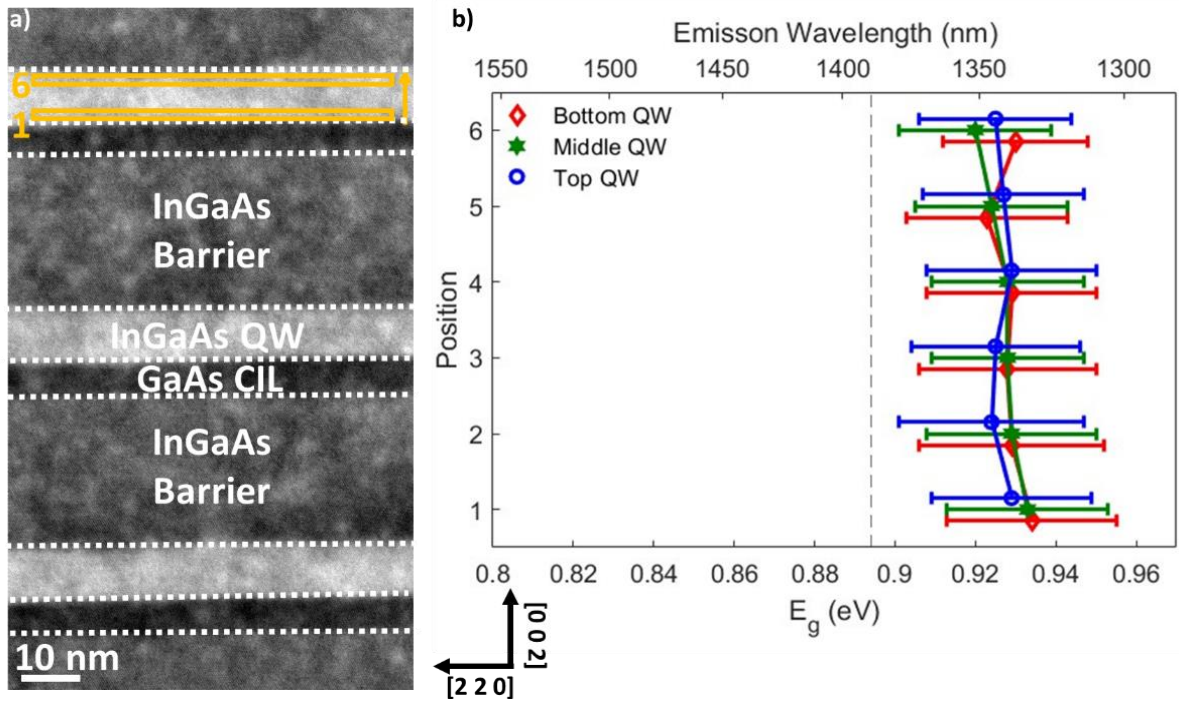

*Figure S9.* $E_g$ *analysis. Representative HAADF-STEM images over all 3 QWs (a), and the corresponding measured $E_g$ corrected for Cerenkov (b): bottom (red diamond), middle (green hexagon) and top QW (blue circle). The $E_g$ was measured horizontally as indicated by the yellow markings for the top QW in a). The secondary axis indicates the corresponding emission wavelength.*

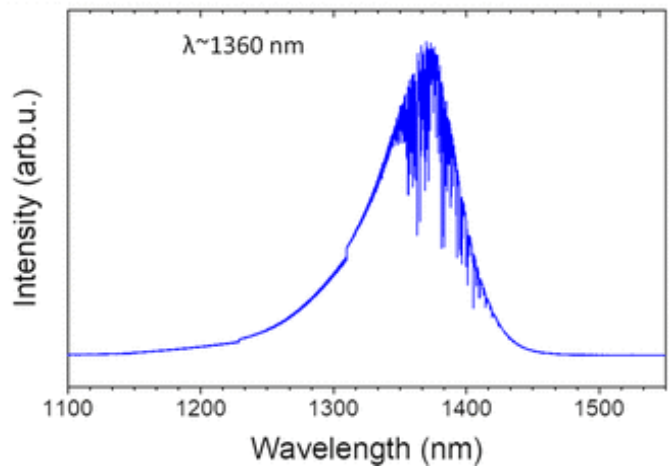

*Figure S10. Photoluminescence spectra of QWs. From* [1].

The lamella thickness was investigated by measuring the relative thickness (t/λ) across the QWs in the sample used for low loss EELS, a different sample was used for EDX. The energy range for the EELS dataset used was from -2.3eV to 18.3eV. This was done using the relative thickness function in Digital Micrograph across all QWs and then taking the t/λ value at each position, identical to those positions in the band gap measurements, Fig. S11. The area of the bottom QW is the thinnest and the area of the top QW is the thickest, ranging from ~0.42 to 0.52 t/λ. A common feature is that the relative thickness increases at the centre of each QW. It should be mentioned that the relative thickness is expected to change with composition (In content), which is likely the case considering the relative thickness profile in Figs. S11 and S2. The mean free path at 60kV for InAs is ~35 nm and ~38 nm for GaAs, with any composition in between having an inelastic mean free path difference of a couple of nms. Moreover, it should be noted that the lowest $E_g$ value is observed for the middle QW, and the highest value for the bottom QW. This is a different order to the thickness changes observed in Fig. S11. Thus, the change in relative thickness for the individual QWs does not seem to affect the changes observed for the individual QWs with respect to the $E_g$ values. Upon examining the $E_g$ measurements, it was determined that the thickness difference in the lamellae is not large enough to affect the measurements.

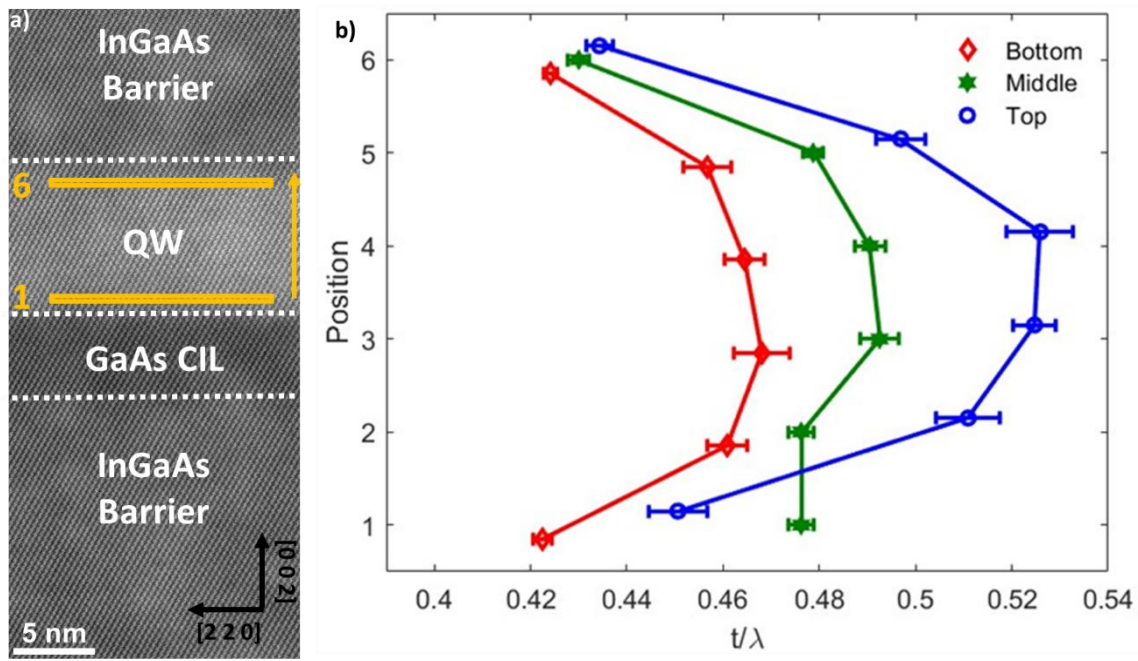

***Figure S11***. *Relative thickness (t/λ) plot at each position for the bottom (red diamond), middle (green hexagon) and top (blue circle) $In_{0.40}Ga_{0.60}As$ QW.*

The Cerenkov corrected $E_g$ and measured In concentration for the middle and top QW are presented in Figure S12. For all QWs, positions 2-5 compare well with the optical $E_g$ with the $E_g$ at position 1 (the GaAs CIL interface) is significantly lower. For position 6 (the $In_{0.13}Ga_{0.87}As$ barrier interface) in the top QW, the measured $E_g$ is significantly lower compared to the optical $E_g$ as was seen in the bottom QW. In contrast, the $E_g$ for the middle QW is within the optical $E_g$ after considering uncertainty. The values and uncertainties for both x in $In_xGa_{1-x}As$ and the $E_g$ in Fig. S12 are presented in Table S4.

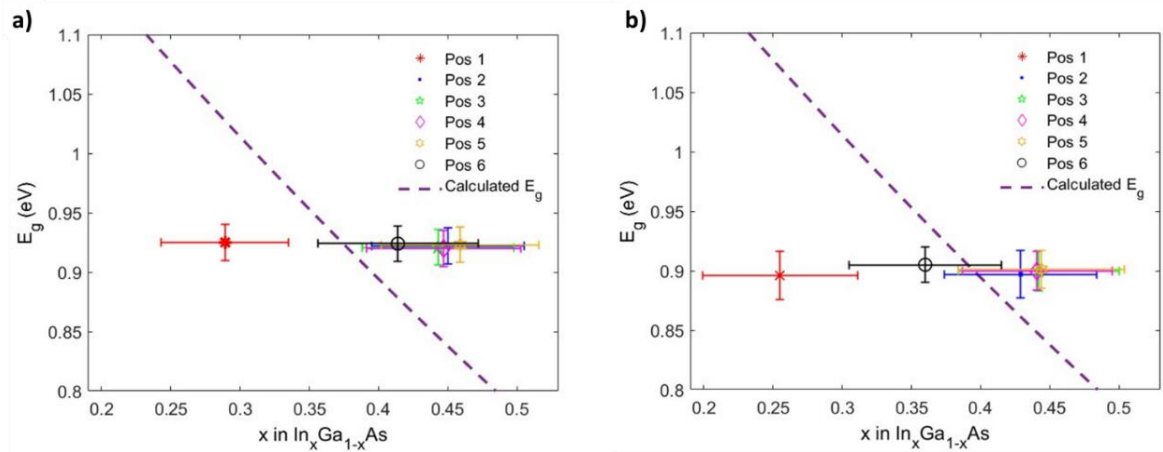

*Figure S12.* Comparison of the calculated and measured Cerenkov corrected $E_g$ values at each position (1-6) for the middle (a) and top QW (b) as function of In mole fraction (x). The calculated $E_g$ as a function of x (dotted purple line) based on Eq. 2 from Nahroy et al .[3]

| QW | Position | Non-Corrected $E_g$ (eV) | Corrected $E_g$ (eV) | x in $In_xGa_{1-x}As$ |
|---|---|---|---|---|
| **Bottom** | 1 | 0.713±0.012 | 0.881±0.019 | 0.316±0.058 |
| | 2 | 0.726±0.015 | 0.866±0.022 | 0.433±0.058 |
| | 3 | 0.708±0.015 | 0.877±0.022 | 0.433±0.055 |
| | 4 | 0.705±0.014 | 0.875±0.021 | 0.437±0.051 |
| | 5 | 0.721±0.011 | 0.884±0.018 | 0.440±0.060 |
| | 6 | 0.745±0.010 | 0.897±0.018 | 0.339±0.054 |
| | Average | 0.720±0.012 | 0.883±0.020 | 0.395±0.081 |
| **Middle** | 1 | 0.794±0.012 | 0.925±0.016 | 0.274±0.044 |
| | 2 | 0.786±0.012 | 0.921±0.016 | 0.439±0.055 |
| | 3 | 0.785±0.011 | 0.920±0.016 | 0.460±0.056 |
| | 4 | 0.785±0.011 | 0.919±0.017 | 0.468±0.055 |
| | 5 | 0.792±0.013 | 0.924±0.017 | 0.462±0.058 |
| | 6 | 0.795±0.014 | 0.926±0.016 | 0.422±0.056 |
| | Average | 0.790±0.012 | 0.922±0.016 | 0.419±0.083 |
| **Top** | 1 | 0.747±0.014 | 0.898±0.020 | 0.262±0.054 |
| | 2 | 0.750±0.014 | 0.900±0.019 | 0.417±0.054 |
| | 3 | 0.757±0.012 | 0.905±0.017 | 0.428±0.056 |
| | 4 | 0.758±0.010 | 0.906±0.016 | 0.439±0.049 |
| | 5 | 0.762±0.008 | 0.908±0.014 | 0.448±0.061 |
| | 6 | 0.767±0.008 | 0.911±0.014 | 0.355±0.054 |
| | Average | 0.757±0.011 | 0.905±0.017 | 0.393±0.081 |

***Table S4.*** *Measured non-corrected and Cerenkov corrected $E_g$ and x in $In_xGa_{1-x}As$ at each position for each $In_{0.40}Ga_{0.60}As$ QW.*

Comparison of uncorrected Cerenkov $E_g$ not corrected for Cerenkov and measured In concentration (Fig. S13) show that even with the inclusion of uncertainty, the measured $E_g$ and In atomic fraction does not fall within the optical $E_g$ from Nahory [3]. This further validates the importance of correcting the Cerenkov radiation effects to ensure an accurate $E_g$ measurement.

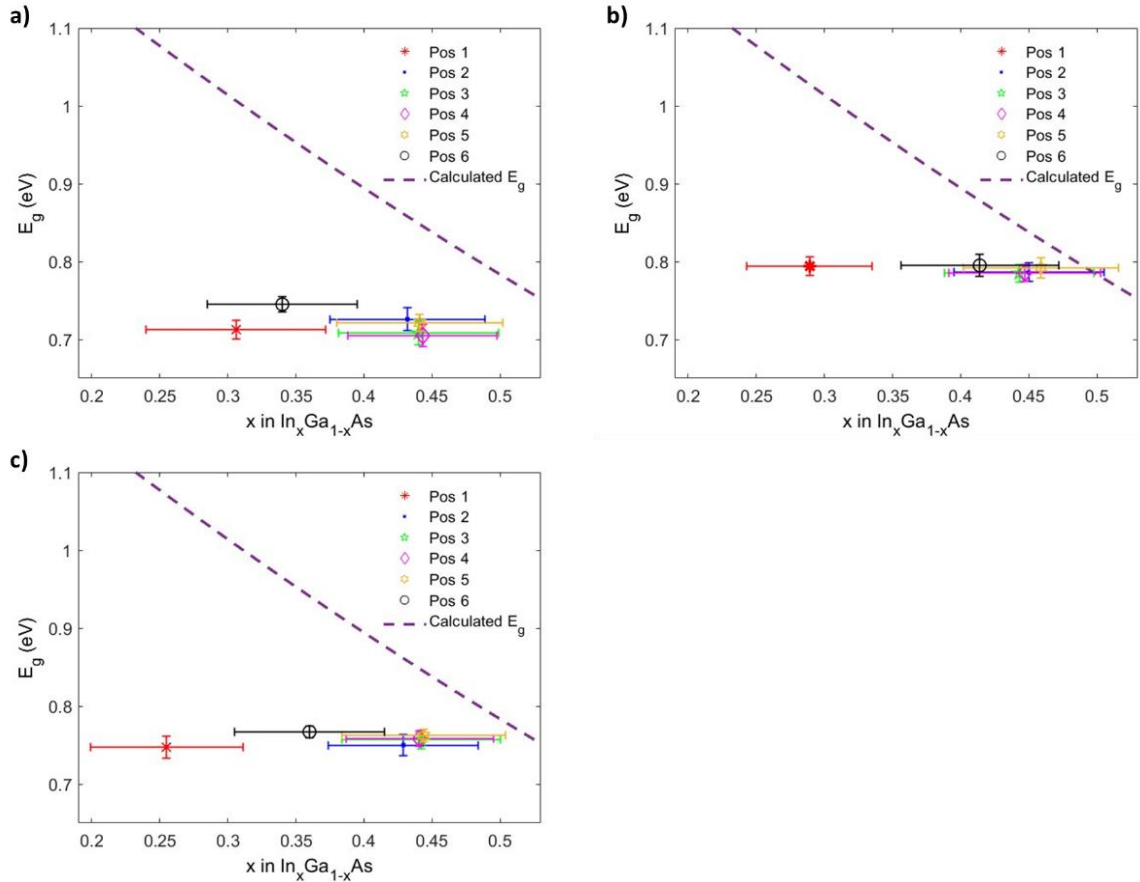

***Figure S13.*** *Comparison of the calculated $E_g$ and measured non-Cerenkov corrected $E_g$ values at each position (1-6) in the bottom (a), middle (b) and top QW (c) as function of In mole fraction (x). The calculated $E_g$ as a function of x (dotted purple line) based on Eq. 2 from Nahroy et al.[44].*

Figure S14 presents the simulated EELS for the outlined concentrations measured. The onset energy from each of the simulated EELS of $In_xGa_{1-x}As$ alloys were extracted by finding the energy where the loss function is significantly greater than zero *i.e* above the x-axis. Initial numbers for the loss function were in the magnitude of $10^{-17}$ which were considered close to zero. It was decided that the $<10^{-5}$ was the smallest number significantly above zero hence the lowest energy was considered to be the onset energy for Fig. 5c in the main manuscript.

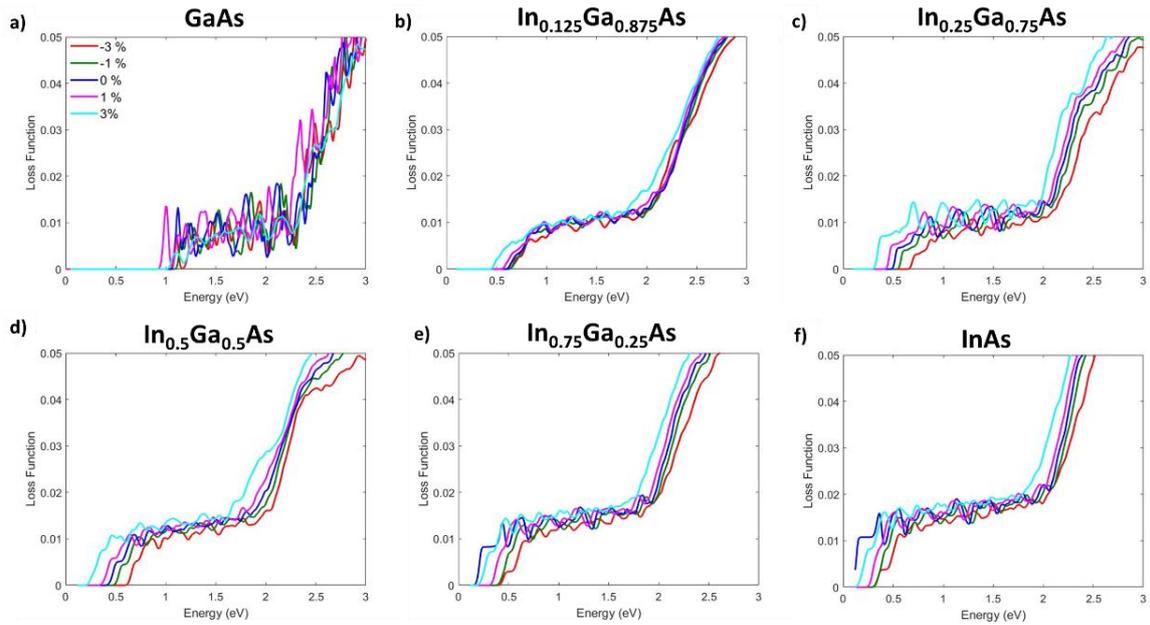

***Figure S14.*** *Simulated EELS of GaAs (a), $In_{0.125}Ga_{0.875}As$ (b), $In_{0.25}Ga_{0.75}As$ (c), $In_{0.50}Ga_{0.50}As$ (d), $In_{0.75}Ga_{0.25}As$ (e) and InAs (f). Coloured lines denote the strain in each of the $In_xGa_{1-x}As$ alloys.*

## References

[1] I. Tångring, et al., *Journal of Crystal Growth* **2007**, 301-302, 971.
[2] M. Arai, et al., *IEEE Journal of Selected Topics in Quantum Electronics* **2013**, 19, 1502207.
[3] D. Wu, et al., *Electronics Letters* **2008**, 44, 474.
[4] A. Benali, et al., *Journal of Crystal Growth* **2022**, 593, 126768.
[5] I. Camps, et al., *Semiconductor Science and Technology* **2002**, 17, 1013.
[6] K. Takemasa, et al., *IEEE Photonics Technology Letters* **1998**, 10, 495.
[7] P. J. Delfyett, in *Encyclopedia of Physical Science and Technology (Third Edition)*, (Ed: R. A. Meyers), Academic Press, New York 2003.
[8] F. Bugge, et al., *physica status solidi c* **2006**, 3, 423.
[9] G. Alahyarizadeh, et al., *International Journal of Modern Physics B* **2015**, 29, 1550081.
[10] D. Saxena, et al., *Nano Letters* **2016**, 16, 5080.
[11] P. A. Chen, et al., *IEEE Journal of Quantum Electronics* **1993**, 29, 2607.
[12] D. S. Sizov, et al., *Applied Physics Express* **2010**, 3, 122101.
[13] T. Uchida, et al., *Electronics Letters* **1994**, 30, 563.
[14] A. E. Zhukov, et al., *Physics of Semiconductor Devices* **2003**, 37, 1119
[15] I. Vurgaftman, et al., *Journal of Applied Physics* **2001**, 89, 5815.
[16] C. P. Kuo, et al., *Journal of Applied Physics* **1985**, 57, 5428.
[17] D. Shiri, et al., *Applied Physics Letters* **2008**, 93, 073114.
[18] P. J. A. Thijs, et al., *Philips Journal of Research* **1995**, 49, 187.
[19] J. Pető, et al., *npj 2D Materials and Applications* **2019**, 3, 39.
[20] S. J. Sweeney, et al., *Journal of Applied Physics* **2019**, 125, 082538.
[21] M. Gal, et al., *Applied Physics Letters* **1988**, 53, 113.
[22] Y. Sun, et al., *Journal of Applied Physics* **2007**, 101, 104503.
[23] P. Makuła, et al., **2018**, 9, 6814
[24] P. V. Sushko, S. A. Chambers, *Scientific Reports* **2020**, 10, 13028.
[25] M. A. Reshchikov, *Journal of Applied Physics* **2021**, 129, 121101.

[26]	R. Erni, N. D. Browning, *Ultramicroscopy* **2007**, 107, 267.
[27]	S. Martí-Sánchez, et al., *Nature Communications* **2022**, 13, 4089.
[28]	O. L. Krivanek, et al., *Ultramicroscopy* **2019**, 203, 60.
[29]	W. Zhan, et al., *Scientific Reports* **2018**, 8, 848.
[30]	W. Zhan, et al., *Nanotechnology* **2017**, 28, 105703.
[31]	D. Keller, et al., *Microscopy and Microanalysis* **2014**, 20, 1246.
[32]	J. R. Jinschek, et al., *Solid State Communications* **2006**, 137, 230.
[33]	J. Wei, et al., *Nano Letters* **2020**, 20, 2530.
[34]	A. Brokkelkamp, et al., *The Journal of Physical Chemistry A* **2022**, 126, 1255.
[35]	M. Pelaez-Fernandez, et al., *Nanomaterials* **2021**, 11, 3218.
[36]	D. V. Beznasyuk, et al., *Physical Review Materials* **2022**, 6, 034602.
[37]	E. E. Mura, et al., *Crystal Growth & Design* **2021**, 21, 2068.
[38]	K. H. Goetz, et al., *Journal of Applied Physics* **1983**, 54, 4543.
[39]	T. Fleischmann, et al., *Journal of Applied Physics* **2001**, 89, 4689.
[40]	T. Grieb, et al., *Ultramicroscopy* **2022**, 238, 113535.
[41]	D. V. Sridhara Rao, et al., *Microscopy and Microanalysis* **2014**, 20, 1262.
[42]	C. S. Granerød, et al., *Ultramicroscopy* **2018**, 184, 39.
[43]	C. A. Gadre, et al., *Nature* **2022**, 606, 292.
[44]	R. E. Nahory, et al., *Applied Physics Letters* **1978**, 33, 659.
[45]	G. H. Olsen, et al., *Journal of Applied Physics* **1978**, 49, 5523.
[46]	B. Mondal, R. Tonner-Zech, *Physica Scripta* **2023**, 98, 065924.
[47]	M. E. Aumer, et al., *Applied Physics Letters* **2000**, 77, 821.
[48]	J. F. Chen, et al., *Journal of Applied Physics* **2000**, 87, 1251.
[49]	J. J. P. Peters, et al., *Ultramicroscopy* **2015**, 157, 91.
[50]	N. Stephen, et al., *Journal of Materials Science* **2023**, 58, 9547.
[51]	H. A. Fonseka, et al., *Nanoscale* **2017**, 9, 13554.
[52]	D. N. Quang, N. H. Tung, *Physical Review B* **2008**, 77, 125335.
[53]	J. Park, et al., *Ultramicroscopy* **2009**, 109, 1183.
[54]	M. Horák, M. Stöger-Pollach, *Ultramicroscopy* **2015**, 157, 73.
[55]	P. Mori-Sánchez, A. J. Cohen, *Phys. Chem. Chem. Phys.* **2014**, 16, 14378.
[56]	J. P. Perdew, et al., *Physical Review Letters* **1982**, 49, 1691.
[57]	K. Lejaeghere, et al., *Science* **2016**, 351, aad3000.
[58]	Y. Ji, et al., *The Journal of Physical Chemistry A* **2022**, 126, 5924.
[59]	M. Oloumi, C. C. Matthai, *Journal of Physics: Condensed Matter* **1991**, 3, 9981.
[60]	J. Tersoff, *Physical Review B* **1984**, 30, 4874.
[61]	V. Dimastrodonato, et al., *Journal of Crystal Growth* **2011**, 315, 119.
[62]	M. Lekstrom, et al., *Journal of Physics: Conference Series* **2008**, 126, 012028.
[63]	C. J. Powell, *NIST* **1976**, 460, 97.
[64]	Geometrical phase analysis software, https://er-c.org/index.php/software/stem-data-analysis/gpa/,
[65]	L. Jones, et al., *Advanced Structural and Chemical Imaging* **2015**, 1, 8.
[66]	D. J. Dunstan, *Journal of Materials Science: Materials in Electronics* **1997**, 8, 337.
[67]	O. L. Krivanek, et al., *Journal of Physics: Conference Series* **2014**, 522, 012023.
[68]	N. Bonnet, et al., *Ultramicroscopy* **1999**, 77, 97.
[69]	K. L. Y. Fung, et al., *Ultramicroscopy* **2020**, 217, 113052.
[70]	S. J. Clark, et al., *Zeitschrift für Kristallographie - Crystalline Materials* **2005**, 220, 567.
[71]	A. P. Bartók, J. R. Yates, *The Journal of Chemical Physics* **2019**, 150, 161101.
[72]	R. J. Nicholls, et al., *Journal of Physics: Conference Series* **2012**, 371, 012062.
[73]	A. J. Morris, et al., *Computer Physics Communications* **2014**, 185, 1477.
[74]	J. R. Yates, et al., *Physical Review B* **2007**, 75, 195121.